\documentclass[%
 reprint,
superscriptaddress,
 amsmath,amssymb,
 aps,prl
]{revtex4-2}

\usepackage{graphicx}
\usepackage{dcolumn}
\usepackage{bm}
\usepackage{xcolor}
\usepackage[utf8]{inputenc}
\usepackage[T1]{fontenc}
\usepackage{mathptmx}
\usepackage{etoolbox}
\usepackage{tikz}
\usepackage{amsmath}
\usetikzlibrary{quantikz}
\usetikzlibrary{tikzmark,calc}
\newdimen\numht
\newdimen\numwd
\newcommand{\RomanNumeralCaps}[1]
    {\MakeUppercase{\romannumeral #1}}

\begin{document}

\preprint{APS/123-QED}

\title{Hybrid Quantum Algorithm for Simulating Real-Time Thermal Correlation Functions} 

\author{Elliot C. Eklund}
\affiliation{Department of Chemistry and Chemical Biology, Cornell University, Ithaca, New York 14853, USA.}
\affiliation{School of Chemistry, University of Sydney, NSW 2006, Australia}
\email{elliot.eklund@sydney.edu.au}
\author{Nandini Ananth}%
\email{ananth@cornell.edu}
\affiliation{Department of Chemistry and Chemical Biology, Cornell University, Ithaca, New York 14853, USA.}
\begin{abstract}
We present a hybrid Path Integral Monte Carlo (hPIMC) algorithm to calculate real-time quantum thermal correlation functions for condensed phase systems. The hPIMC algorithm leverages the successes of classical PIMC as a computational tool for high-dimensional system studies by exactly simulating dissipation using the Feynman-Vernon influence functional on a classical computer. We achieve a quantum speed-up over the classical algorithm by computing short-time matrix elements of the quantum propagator on a quantum computer. We show that the component of imaginary-time evolution can be performed accurately using the recently developed Probabilistic Imaginary-Time Evolution (PITE) algorithm, and we introduce a novel low-depth circuit for approximate real-time evolution under the kinetic energy operator using a Discrete Variable Representation (DVR). We test the accuracy of the approximation by computing the position-position thermal correlation function of a proton transfer reaction.
\end{abstract}

\maketitle
\textit{Introduction.---} 
Noisy intermediate-scale quantum (NISQ) computers are characterized by small qubit counts and short coherence times~\cite{preskill2018}. Algorithms designed for NISQ era hardware must therefore operate within the constraints of limited qubits and low-depth circuits. One way this achieved is by using so-called hybrid quantum algorithms in which inexpensive subroutines are performed classically while expensive, quantum-in-nature calculations are off-loaded to a quantum computer. The most thoroughly explored approaches to hybrid algorithm design are variational quantum algorithms (VQAs), which are inspired by classical variational optimization~\cite{mcclean2016,miessen2021,ollitrault2020,ollitrault2021, yung2014,wang2019,peruzzo2014,lee2022,yao2021,xing2023}. Although VQAs succeed in achieving shallow quantum circuits, they often scale poorly with system size due to the emergence of barren plateaus, with their performance being limited by the expressiveness of their variational ansatz~\cite{wang2021,sack2022,cerezo2021}.

Recently, Layden \textit{et al.}~proposed a hybrid algorithm based on Markov chain Monte Carlo and Metropolis-Hasting acceptance/rejection criteria, and demonstrated that it achieves a quantum speedup over classical methods for sampling the Boltzmann distribution of an Ising spin model~\cite{layden2023}. A related method, used to compute dynamical quantities, is Path Integral Monte Carlo (PIMC)~\cite{bernu2002}. In this paper, we propose the first hybrid PIMC (hPIMC) algorithm and show that it achieves a quantum speedup in the calculation of real-time thermal correlation functions (TCFs) for open quantum systems. Importantly, like Layden's algorithm, hPIMC does not rely on variational techniques and therefore offers a new avenue for exploring chemical dynamics with hybrid computing.

Open quantum systems typically comprise a quantum mechanical system that interacts with and dissipates energy to an external environment that can, in turn, drive the system dynamics. Classical PIMC methods often involve computing short-time matrix elements of the system time-evolution operator (real-time in the case of zero temperature and complex-time for finite temperature simulations) by diagonalizing the Hamiltonian in a first-quantized representation for an arbitrary potential energy surface, and then incorporating the effects of the environment exactly through the Feynman-Vernon influence functional~\cite{makri1998,topaler1992,topaler1993-1,1topaler1993-2,topaler1994,feynman1982}. Although classical PIMC approaches have been used for relatively high dimensional simulations, typically this corresponds to a significant increase in the number of environment degrees of freedom while the system remains low-dimensional, with only a couple continuous degrees of freedom representing the limit beyond which calculation of the system time-evolution operator (TEO) matrix elements become unfeasible. 

The central ideas of the hPIMC approach are to (i) use a quantum computer to push the boundaries of the system size limitations and (ii) leverage the fact that in the path integral framework, only short-time TEO matrix elements are required, eliminating cumulative errors on the quantum computer due to successive applications of short-time quantum propagation steps. We use these ideas to construct a hybrid Monte Carlo algorithm where operations such as sampling random paths and evaluating acceptance/rejection criteria are carried out classically, while the expensive task of computing TEO matrix elements is performed by a low-depth quantum circuit.

In what follows, we introduce hPIMC for quantum systems in an arbitrary representation and show that it achieves a quantum speedup compared with classical PIMC for computing real-time TCFs. We then adapt the hPIMC algorithm to condensed phase systems modeled by the Caldeira-Leggett (CL) Hamiltonian in position space and introduce an influence functional to capture dissipative effects. We also use the PITE algorithm~\cite{kosugi2022,nishi2023} for imaginary-time evolution of the TEO matrix elements necessary for systems at finite temperature initially in thermal equilibrium. We further propose a new approach to approximate system propagation under the kinetic energy operator using a discrete variable representation (DVR)~\cite{light2000,littlejohn2002,colbert1992} and prove an algorithm for doing so. We test the accuracy of this approximation with classical simulations of the position TCF of a proton transfer reaction.

\begin{sloppy}
\textit{Theory.---}
We start with a symmetrized real-time TCF between two operators, $A$ and $B$, at time $t$ for a system at equilibrium, 
\begin{equation} \label{thermal-corr-exact}
    C_{AB}(t) =\frac{1}{Z} \operatorname{Tr}\left(U^{\dag}(t_c) A U(t_c) B \right),
\end{equation}
where $U(t_c) = e^{-i H t_{c}}$ is the complex-time evolution operator, $H$ is the system Hamiltonian, and  $t_{c} = t - i \beta /2$ where $\beta = 1/k_B T$, $k_B$ is the Boltzmann constant, and $T$ is temperature. Throughout, we work in atomic units so that $\hbar = k_{B} = 1$. In a PIMC calculation, the partition function, $Z$, can be computed through a separate normalization calculation \cite{topaler1994,1topaler1993-2}, and we do not explicitly address methods of computing it here.
\end{sloppy}

In most cases, finding an exact representation of $U(t_c)$ is not possible, and approximations are used instead. Let $\tilde{U}(t_c)$ be an approximation to $U(t_c)$ such that the error is bound by
\begin{equation}
    \Vert U(t_c) - \tilde{U}(t_c)\Vert_1 \leq \epsilon_{U},
\end{equation}
where $\Vert \cdot \Vert_1$ is the matrix 1-norm or trace norm. Further, let $\Tilde{C}_{AB}(t)$ be an approximation to ${C}_{AB}(t)$ in which $\Tilde{U}(t_{c})$ is used in place of $U(t_c)$. We show in Supplemental Material, Sec. \RomanNumeralCaps{1} that the error, $\epsilon_{C} = |C_{AB}(t) - \Tilde{C}_{AB}(t)|$, made from approximating $C_{AB}(t)$ is on the order,
\begin{equation} \label{corr-error}
    \epsilon_{C} \leq O\left(\Vert A \Vert_1 \Vert B\Vert_1 \Vert e^{-\beta H /2}
    \Vert_{1} \epsilon_{U}/Z\right).
\end{equation}

Within the path integral framework $U(t_c)$ is discretized into a product of $N$ short-time system TEOs
\begin{equation} \label{therm-corr-prod}
\tilde{C}_{AB}(t) = \frac{1}{Z}\operatorname{Tr}\left(\left(\prod_{k=1}^{N}\tilde{U}^{\dag}(\Delta t_c)\right) 
A \left(\prod_{k=1}^{N} \tilde{U}(\Delta t_c) \right) B \right),
\end{equation}
where $N$ is the number of Trotter steps, and $\Delta t_c = t_c/N$. In order to evaluate $\tilde{C}_{AB}(t)$, we insert copies of identity using a complete basis, $\{\phi_{j}\}$, into Eq.~(\ref{therm-corr-prod}); the choice of basis here is arbitrary. The resulting expression for the TCF is,
\begin{align} \label{corr-mc}
\tilde{C}_{AB}&(t) = \frac{1}{Z} \sum_{J} \left(\prod_{k=N+2}^{2N+1}  
\langle \phi_{j_{k+1}}|\Tilde{U}^{\dag}(\Delta t_c)|\phi_{j_k}\rangle \right)\nonumber\\
&  \langle \phi_{j_{N+2}} | A | \phi_{j_{N+1}}\rangle \left(\prod_{k=1}^{N}  
\langle \phi_{j_{k+1}} |\Tilde{U}(\Delta t_c)| \phi_{j_k} \rangle  \right) 
\langle \phi_{j_1}|B|\phi_{j_0}\rangle
\end{align}
where $J=(j_{0},\dots,j_{2N+2})$ is a set of indices and at each Trotter step, indexed by $k$, we sum over the entire basis set. Note that $\phi_{j_0}=\phi_{j_{2N+2}}$ due to the trace. We sample path space in $\{\phi_{j}\}$ by defining the product
\begin{equation} \label{importance_samp}
    \Theta(\boldsymbol{\phi}) = \left(\prod_{k=N+2}^{2N+1}  
    \langle \phi_{j_{k+1}}|\Tilde{U}^{\dag}(\Delta t_c)|\phi_{j_k}\rangle \!\! \right)\!\!\!\!  
    \left(\prod_{k=1}^{N}  \langle \phi_{j_{k+1}} |\Tilde{U}(\Delta t_c)| \phi_{j_{k}} \rangle \! \!\right)
\end{equation}
and normalization factor
\begin{equation} \label{normalization}
    F = \sum_{J} \vert \Theta(\boldsymbol{\phi}) \vert,
\end{equation}
where $\boldsymbol{\phi} = (\phi_{j_1},\dots,\phi_{j_{2N+2}})$. This allows us to express Eq.~(\ref{corr-mc}) in the conventional PIMC form of an expectation value over a distribution, $W$, which we use for importance sampling:
\begin{align} \label{expect_value}
    \tilde{C}_{AB}(t) &= \frac{F}{Z} \left\langle \langle \phi_{j_{N+2}} |A| \phi_{j_{N+1}}\rangle 
    \langle \phi_{j_1}|B|\phi_{j_0}\rangle \mathrm{sgn} \Theta(\boldsymbol{\phi})  
    \right\rangle_{W}
\end{align}
where $W = \Theta/F$. We note that $F$ can be computed concurrently with $\tilde{C}_{AB}(t)$.

Evaluating $\tilde{C}_{AB}(t)$ with Monte Carlo introduces an additional source of error, $\epsilon_{MC}$, which is dependent on the Monte Carlo sampling rate of convergence. In general, for $M$ Monte Carlo iterations, this error is given by
\begin{equation} \label{mc-error}
    \epsilon_{MC} = O\left(1/\sqrt{M} \right).
\end{equation}
We note that although Eq.~(\ref{mc-error}) holds in the asymptotic limit, the true rate of convergence is more complicated and not easy to express since for small systems with no significant dissipation, the well-established sign problem in quantum dynamics must be considered~\cite{caratzoulas1996}.

In order to tie together both sources of error, we let the total error of computing Eq.~(\ref{thermal-corr-exact}) be divided equally between contributions from $\epsilon_{C}$ and $\epsilon_{MC}$ so that
\begin{equation}
    \epsilon = \epsilon_C /2 + \epsilon_{MC}/2,
\end{equation}
where $\epsilon$ is the total error. To ensure $\epsilon_{MC}/2$ accuracy from Monte Carlo, we require at least $M = O(1/\epsilon_{MC}^2)$ Monte Carlo iterations. We would also like to find a way to relate $\epsilon_C$ to the number of Trotter steps, however, this will depend on the choice of approximation used to implement $\Tilde{U}$. For example, when the Hamiltonian has the form $H = H_1 + H_2$ with non-commuting $H_1$ and $H_2$, $U(t_c)$ can be approximated with a $2k$th-order Suzuki product ~\cite{berry2007}. In this case, to ensure $\epsilon_{C}/2$ accuracy from approximating $U(t_c)$ as $\tilde{U}(t_c)$, the number of Trotter steps must be at least
\begin{equation}
    N = O\left(\left(\Vert A \Vert_1 \Vert B\Vert_1 \Vert e^{-\beta H /2}\Vert_{1} \Omega^{2k+1} |t_c|^{2k+1} 
     /Z \epsilon_C \right)^{1/2k} \right),
\end{equation}
where
\begin{equation}
\Omega = \mathrm{max}\{\Vert H_1 \Vert, \Vert H_2\Vert\}.
\end{equation}

We now estimate the runtime and space complexity of computing $\tilde{C}_{AB}(t)$ using classical PIMC and hPIMC. In either case, within each Monte Carlo iteration, two random strings of matrix elements, $\{\langle \phi_{j_{k+1}}|\Tilde{U}(\Delta t_c)|\phi_{j_{k}}\rangle\}$ and $\{\langle \phi_{j_{k+1}}|\Tilde{U}^{\dag}(\Delta t_c)|\phi_{j_k}\rangle\}$, referred to as the forward and backward paths, respectively, are sampled and the TCF estimators are computed. For classical PIMC, the TEO matrix elements are found by diagonalizing the Hamiltonian expressed in a chosen finite basis. For a $d$-dimensional system and using a basis set of size $P$, the one-time cost of diagonalization is $O(P^{3d})$ and $O(P^d \times P^d)$ space in memory is required to store the result. We assume that the cost to compute and store $\langle \phi_{j_{k+1}}|A|\phi_{j_k}\rangle$ and $\langle \phi_{j_{k+1}}|B|\phi_{j_k}\rangle$ is at most on the same order of $\tilde{U}(\Delta t_c)$. For every Monte Carlo iteration, $O(N)$ elementary operations are needed to compute Eq.~(\ref{therm-corr-prod}) for given forward and backward paths. The total cost for classical PIMC is therefore
\begin{align}
    &\mathrm{PIMC\,\, Runtime} = O(M N + P^{3d})\\
    &\mathrm{PIMC\,\, Space} = O(P^d \times P^d)\nonumber 
\end{align}

For hPIMC we propose to compute each TEO matrix element individually on the quantum computer, thus avoiding the need to diagonalize $\Tilde{U}(\Delta t_c)$. We therefore trade the exponential cost of diagonalization for $O(M N)$ calls to a quantum oracle that implements $\langle \phi_{j_{k+1}}|\Tilde{U}(\Delta t_c)|\phi_{j_{k}}\rangle$. The efficiency of hPIMC therefore depends on the efficiency of the oracle. However, for many important chemical systems, efficient implementations of $\tilde{U}(\Delta t_{c})$ in terms of both runtime and space complexity are known~\cite{coa2019,babbush2018,su2021, babbush2019,lee2022,childs2022,kassal2008,ollitrault2020}. We assume that $A$ and $B$ can be computed classically with negligible cost. To represent each $|\phi_{j_k}\rangle$ on a quantum computer, $d$ quantum registers, each consisting of $n=\log_2(P)$ qubits, are needed. The total cost of hPIMC is therefore
\begin{align} \label{complexities}
    &\mathrm{hPIMC \,\, Runtime} = O(M N Q(\tilde{U})) \\
    &\mathrm{hPIMC \,\, Space} = O(d n) + \mathrm{Anc.}(\Tilde{U}) \nonumber
\end{align}
where $Q(\tilde{U})$ and $\mathrm{Anc.}(\tilde{U})$ are the cost and number of ancillary qubits, respectively, needed to call an oracle that implements $\tilde{U}(\Delta t_c)$. From this analysis, we see that hPIMC scales exponentially better in terms of both space and runtime complexity compared to classical PIMC. Additionally, because each TEO matrix element involves only a single application of $\tilde{U}(\Delta t_{c})$, the quantum circuit is low depth and cumulative error from computing a series of $\tilde{U}(\Delta t_{c})$ on a noisy quantum computer is avoided.

We note that we can further improve the performance of hPIMC by storing TEO matrix elements in a classical look-up table and querying it each Monte Carlo iteration. Then, the distribution between classical and quantum resources becomes increasingly more classical as the calculation progresses and the table fills up. Further, employing problem-specific sampling techniques will likely allow us to target the most relevant Hilbert subspace, keeping the table from becoming exponentially large.

Finally, it is clear that hPIMC offers a significant advantage over VQAs in that for a fixed number of qubits, the error can be systematically reduced without increasing the depth of the quantum circuit by running more Monte Carlo iterations, as shown in Eq.~(\ref{mc-error}), or increasing $N$. This is not the case for VQAs: for a fixed number of qubits, and assuming that the classical optimizer has found the best solution, the error of a VQA is related to the expressiveness of the variational ansatz. Increasing the expressiveness, and thereby reducing the error, requires adding more gates to the variational ansatz, which increases circuit depth.

\textit{Open Quantum Systems.---} Much of the success of classical PIMC has been in the context of dissipative systems at finite temperature where the rapid damping of quantum system oscillations by coupling to a large number of environment degrees of freedom significantly reduces the dynamical sign problem.  This motivates us to adapt hPIMC for condensed phase dynamics described by the CL Hamiltonian using a position space representation. Specifically, the CL Hamiltonian describes a primary system consisting of $\eta$ modes in $d$ physical dimensions bilinearly coupled to an isotropic bath approximated by $f$ harmonic modes,
\begin{equation} \label{HCL}
    H = H_{s} + H_{b},
\end{equation}
where $H_{s}$ is the system sub-Hamiltonian shifted along the adiabatic path and  $H_{b}$ is the bath sub-Hamiltonian modified to include the system-bath coupling term~\cite{topaler1993-1}.  It has been shown that for this choice of sub-Hamiltonians the error made by approximating the TEO as 
\begin{equation} \label{splitOP}
    e^{-i H \Delta t_c} \approx e^{-i H_{b} \Delta t_c/2}  e^{-i H_{s} \Delta t_c}  e^{-i H_{b} \Delta t_c/2} 
\end{equation}
goes as $O( \Lambda \vert \Delta t_c \vert^3)$, where $\Lambda$ depends on the adiabaticity of the bath and goes to zero in the limit that the bath responds instantaneously to the system~\cite{makri1998}.

Another reason for expressing the CL Hamiltonian in this way is that in a position space representation, all terms involving contributions from the bath can be collected and integrated analytically, yielding the Feynman-Vernon influence functional~\cite{thomas2016}. Assuming $A$ and $B$ are functions only of position, the TCF of Eq.~(\ref{HCL}) becomes
\begin{equation} \label{quapi}
\tilde{C}_{A B}(t) = \frac{1}{Z} \int d\mathbf{x} \, P_{b}(\mathbf{x}_{b}) A\left(\mathbf{x}_{N}\right) P_{f}(\mathbf{x}_{f}) B(\mathbf{x}_{0}) I(\mathbf{x}),
\end{equation}
where $I(\mathbf{x})$ is the influence functional and we define the forward path, 
$\mathbf{x}_{f} = (\mathbf{x}_{0},\dots,\mathbf{x}_{N})$, the backward path, 
$\mathbf{x}_{b} = (\mathbf{x}_{N},\dots,\mathbf{x}_{2N})$, and 
$\mathbf{x} = (\mathbf{x}_{f}, \mathbf{x}_{b})$. 
We also define the product of forward and backward path matrix elements
\begin{equation}
    P_{f}(\mathbf{x}_{f}) = \prod_{k=0}^{N-1} 
    \langle \mathbf{x}_{k+1} |e^{-i H_{s} \Delta t_c}| \mathbf{x}_{k} \rangle
\end{equation}
and
\begin{equation}
    P_{b}(\mathbf{x}_{b}) = \prod_{k=N}^{2N-1 } \langle \mathbf{x}_{k+1} |e^{i H_{s} \Delta t_c^{*} }| \mathbf{x}_{k} \rangle,
\end{equation}
respectively. Monte Carlo importance sampling is performed using $P_{f}$ and $P_{b}$ in Eqs.~(\ref{importance_samp}-\ref{expect_value}). Before integrating out the bath degrees of freedom, the integral in Eq.~(\ref{quapi}) is $2Nd(\eta + f)$ dimensional. To accurately model the bath using Eq.~(\ref{HCL}) we typically have $f \gg \eta$. Using the influence functional reduces the dimensionality of the integral to $2N d \eta$, which is the dimensionality of the same system if it were in the gas phase. The influence functional can be computed on a classical computer with negligible cost.

We can approximate the continuous integrals in Eq.~(\ref{quapi}) as discrete sums over a finite width grids of evenly spaced grid points. In one dimension, let $L$ be the total length of the grid and $D$ the number of grid points. Any position on the grid is given by $x_{q} = -\frac{L}{2} + q \Delta x$, where $\Delta x = L/D$, $q \in G$, and $G$ is the set of integers $G=[0,D-1]$. We represent the grid on a quantum computer by mapping $q$ to the computational basis:
\begin{equation}
    \vert q \rangle = \vert b_{q_{n-1}} \rangle \otimes 
    \vert b_{q_{n-2}}\rangle \otimes \dots \vert b_{q_{0}}\rangle,
\end{equation}
where $b_{q_{i}}$ are the binary expansion coefficients of $q$ and $n = \lceil\log_2 D \rceil$. Here, $\vert q \rangle$ represents a single quantum register. For $\eta d$ dimensions, each position coordinate is mapped to its own quantum register and the overall state is the direct product of all $\eta d$ registers. The constant offset, $-L/2$, and grid spacing, $\Delta x$, are absorbed into the implementation of the potential energy~\cite{beneti2008}. Because we are mapping positions on a grid directly to the computational basis, each register can be efficiently initialized with a single layer of Pauli-X gates.

We compute the real and imaginary parts of $\langle \mathbf{x}_{k+1}|\tilde{U}(\Delta t_c) |\mathbf{x}_{k} \rangle$, where now $\tilde{U}(\Delta t_c) = e^{-i H_{s} \Delta t_c}$, on a quantum computer using a standard Hadamard test~\cite{aharonov2009}. In doing so, we split the TEO into a product of real and imaginary-time TEOs, 
\begin{equation} \label{h-split-time}
e^{-i H_{s} \Delta t_{c} } = e^{- i H_{s}\Delta t} 
e^{- \frac{\Delta \beta}{2} H_{s}},
\end{equation}
where $\Delta t = t/N$ and $\Delta \beta= \beta/N$. We perform imaginary-time evolution using the PITE algorithm~\cite{kosugi2022,nishi2023}. For an initial state $|\Psi\rangle$ and a single ancilla qubit, PITE takes as input a circuit that executes real-time evolution, $e^{-i  H_{s}\tau}$, and approximates evolution in imaginary-time by the unitary $U_{PITE}$ such that
\begin{align}
&U_{PITE} |\psi\rangle |0\rangle \rightarrow   
m_0(1- H_{s} \tau)|\psi\rangle \otimes|0\rangle  \nonumber\\
&+\left(\sqrt{1-m_0^2}+\frac{m_0^2}{\sqrt{1-m_0^2}} H_{s} 
\tau\right)|\psi\rangle \otimes|1\rangle  + O\left(\tau^2\right),
\end{align}
where $ \tau = s_{1} \Delta \beta/2$, $s_{1} = m_{0}/\sqrt{1 - m_{0}^{2}}$, and $m_{0}$ is a real parameter. When the ancillary qubit is measured in the state $|0\rangle$, $|\psi \rangle$ is evolved in imaginary-time by a first-order approximation to $e^{- \frac{\Delta \beta}{2} H_{s}}$. Note that the probability of successfully measuring $|0\rangle$ increases for smaller values of $\tau$. PITE was originally designed for finding the ground state of an input Hamiltonian where successive applications of the circuit are needed to reach larger $\tau$, each of which lowers the probability of success. However, in hPIMC, $\tau$ is inherently small to ensure that Eq.~(\ref{splitOP}) is a good approximation. Thus we only require a single application of PITE per TEO matrix element which means the additional overhead from calling PITE until a successful outcome is measured is minimal.

For evolution in real-time, we approximate the TEO using a second-order Trotter splitting,~\cite{trotter1959}
\begin{equation} \label{real-time-trot}
    e^{-iH_{s}\Delta t} \approx e^{-iV\Delta t/2} 
    e^{-iT\Delta t}e^{-iV\Delta t}.
\end{equation}
In the position basis, $V$ is diagonal and $e^{-iV\Delta t/2}$ can be constructed efficiently if there exists a classical algorithm that efficiently implements $e^{-iV\Delta t/2}$~\cite{aharonov2003}. For example, quadratic potentials with linear coupling can be constructed with $O(\eta d n^{2})$ one- and 
two-qubit gates.~\cite{beneti2008}

Although the CL Hamiltonian described here includes only a single electronic state potential energy surface, we note that hPIMC can easily be extended to nonadiabatic systems that employ a diabatic representation for multiple potential energy surfaces following the work in~\cite{ollitrault2020}. We expect hPIMC to provide a useful tool for studies of condensed phase nonadiabatic systems in which the electronic state transitions are strongly coupled to multiple vibrational degrees of freedom, but save this investigation for future work.

\textit{DVR Approximation.---} 
We propose to evaluate time evolution under the kinetic energy operator with a discrete variable representation (DVR) of $T$. The DVR of the kinetic energy operator expressed using a sinc basis set is given by~\cite{light2000}
\begin{equation} \label{high-d-dvr}
    T^{DVR} = \sum_{\alpha=1}^{\eta} T_{\alpha}^{DVR},
\end{equation}
where
\begin{equation}
    T_{\alpha}^{DVR} = I_{1}^{\otimes d}\otimes \dots 
    I_{\alpha-1}^{\otimes d} \otimes DVR^{\otimes d}_{\alpha} 
    \otimes  I_{\alpha+1}^{\otimes d} \otimes \dots 
    \otimes I_{\eta}^{\otimes d}.
\end{equation}
Here, $I^{\otimes d}$ is a tensor product of $d$ identity matrices and the DVR of the $\alpha$th system mode in one dimension is defined by the matrix elements
\begin{equation}\label{dvr-elements}
DVR_{\alpha,ij} = \frac{(-1)^{i-j}}{2m_{\alpha} (\Delta x)^2} \begin{cases}
\frac{\pi^2}{3}, & i=j\\
\frac{2}{(i-j)^2}, & i\neq j
\end{cases}
\end{equation}
where $m_{\alpha}$ is the mass of the $\alpha$th system mode. In the limit where $\Delta x$ goes to zero, this approximation is exact. 

Ordinarily, one diagonalizes the DVR Hamiltonian, $H^{DVR} = T^{DVR} + V$, and performs dynamics in the basis of DVR eigenvectors. However, in hPIMC, we avoid diagonalizing $H^{DVR}$ by approximating evolution under $T$ as $T^{DVR}$. That is, 
\begin{equation} \label{sim-t}
    e^{-iT \Delta t} \approx e^{-i T^{DVR} \Delta t} =\prod_{\alpha = 1}^{\eta}  e^{-i T^{DVR}_{\alpha}\Delta t}.
\end{equation}
Since DVR is not typically used this way, we present results in the next section that validate this approach.

In general, we cannot efficiently implement $e^{-i DVR_{\alpha} \Delta t }$ on a quantum computer because $DVR_{\alpha}$ is dense. At the same time, we can see from Eq.~(\ref{dvr-elements}) that the matrix elements decrease in magnitude moving away from the main diagonal, and should eventually become small enough to safely neglect. In Supplemental Material, Sec.~\RomanNumeralCaps{2}, we prove that this is indeed the case and find that for a given error tolerance, $\delta$, $DVR_{\alpha}$ can be made sparse by neglecting all diagonals, $\nu$, where $\nu \geq \ell$, and $\ell$ is given by
\begin{equation} \label{diagonal-bound}
    \ell = \left\lfloor \frac{(2 K)^{2/3}}{\delta^{2/3}}\right\rfloor,
\end{equation}
with $K= 1/m_{\alpha} (\Delta x)^2$. Here, $\nu=1$ is the main diagonal, $\nu=2$ are the adjacent upper and lower diagonals, and so on. Formally, this expression results from a lower bound on error; however, we show in Supplemental Material, Sec.~\RomanNumeralCaps{2} that this bound is tight. 

\begin{figure*} \label{fig:corr_fxns}
    \centering
\includegraphics[width=0.7\textwidth]{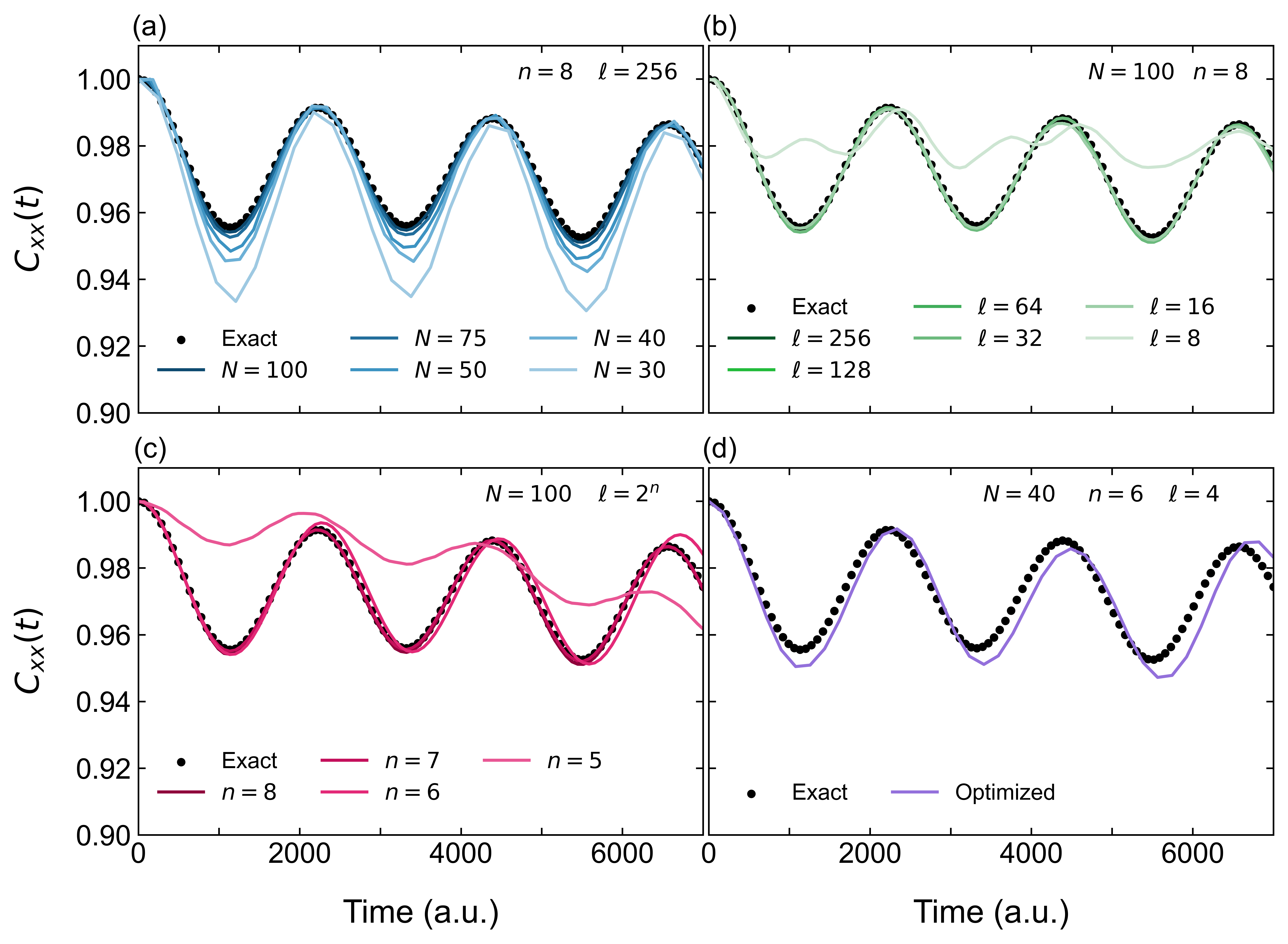}
    \caption{TCF of symmetric double-well potential. We evaluate real-time TEO matrix elements using a second-order Trotter splitting about $T$ and $V$, and QFT for evolution under the kinetic energy operator. For imaginary-time evolution, we use a first-order Trotter splitting about $T$ and $V$, and  Eq.~(\ref{final-dvr}) for evolution under the kinetic energy operator. For each plot, $m=1836$~a.u., $1/\beta = 350\,\mathrm{K}$, $\omega_{b}=500 \,\mathrm{cm}^{-1}$, $V_{0}=1500\, \mathrm{cm}^{-1}$, and $L=30$~a.u. In (a) we vary the number of Trotter steps $N$, (b) we vary the number of grid points, $D=2^n$, and (c) we further approximate the DVR by neglecting all diagonals $\nu \geq \ell$. In (d), we plot the TCF using optimized parameters so that the resource requirements are minimized while maintaining reasonable error. Each plot is normalized by the magnitude of its maximum value.}
    \label{corr_fxns}
\end{figure*}

We can then simulate time evolution under $DVR_{\alpha}$ using a first-order Trotter splitting about the diagonals of $DVR_{\alpha}$:
\begin{equation}\label{dvr-trotter}
    e^{-i DVR_{\alpha} \Delta t} \approx \prod_{\nu=1}^{\ell} 
    e^{-i \, \mathrm{diag}\left( DVR_{\alpha},\nu\right) \Delta t},
\end{equation}
where $\mathrm{diag}(\cdot,\nu)$ gives the input matrix with all elements neglected except those on the $\nu$th diagonals. Each matrix $\mathrm{diag}\left( DVR_{\alpha},\nu\right)$ is 2-sparse and in Supplemental Material, Sec.~\RomanNumeralCaps{3} we prove that each of these 2-sparse matrices can be decomposed into a sum of at most two 1-sparse matrices. Our method yields a factor of 12 improvement compared to the results of~\cite{berry2007} for a 2-sparse matrix. Using a first-order Trotter splitting about each 1-sparse matrix yields
\begin{equation}\label{final-dvr}
    e^{-i DVR_{\alpha} \Delta t} \approx \prod_{\nu=1}^{\ell} 
    \prod_{\sigma = 1}^{2} e^{-i \, 
    \mathrm{diag}\left( DVR_{\alpha},\nu\right)_{\sigma} \Delta t},
\end{equation}
where $\sigma$ indexes the 1-sparse matrices resulting from the decomposition. From here, each 1-sparse matrix can be directly simulated with just two black-box queries to $\mathrm{diag}(DVR_{\alpha},\nu)_\sigma $~\cite{aharonov2003,childs2003}. In total, Eq.~(\ref{sim-t}) can be simulated with at most $4 \eta d \ell$ black-box queries.

\textit{Results and Conclusion.---}
Our proposed method to simulate the system TEO involves several approximations, and although the error from some of these approximations can be analyzed individually, rigorously studying the cumulative error is not straightforward. Instead, we investigate the resulting error by computing the position-position TCF of a proton transfer reaction using the approach outlined above and compare the results to the exact ones obtained by diagonalization. We model the donor and acceptor sites of the reaction using a standard symmetric double-well potential:
\begin{equation}
    V(x) = -\frac{1}{2} m \omega_{b}^2 x^2 + \frac{m^2\omega_b^4}{16 V_{0}}x^{4}.
\end{equation}
Real-time evolution is approximated with a second-order Trotter splitting, as in Eq.~(\ref{real-time-trot}), and imaginary-time evolution is approximated with a first-order Trotter splitting to replicate PITE. We evaluate the TCF using quadrature to avoid Monte Carlo error. All results are obtained with a classical computer.

Initially, we used Eq.~(\ref{final-dvr}) to approximate evolution under the kinetic energy operator for both real- and imaginary-time. However, we found that the value of $N$ needed to achieve reasonable error was too large for PIMC methods. Instead, we found that using Eq.~(\ref{final-dvr}) for imaginary-time only, and using a standard Quantum Fourier Transform (QFT) approach to diagonalize the kinetic energy operator in real-time yielded better results. The results presented in Fig.~\ref{corr_fxns} were obtained using this approach. Interestingly, we also found that using QFT for both real- and imaginary-time evolution yielded worse results than using Eq.~(\ref{final-dvr}) for both, and that the results do not systematically improve with increasing $N$ as one would expect.  

In Fig.~\ref{corr_fxns}, panel $\mathrm{(a)}$, we see that for sufficiently large $N$ the approximate results match the exact results. As $N$ decreases, the error from Trotter splitting increases, and we see the amplitude of the TCF begins to deteriorate, although the frequency of the oscillations remains accurate. In panel $\mathrm{(b)}$, we compare approximate and exact results for different numbers of grid points, $D=2^n$. For $n=8$ and $7$, the difference between them is indistinguishable. At $n=6$, the approximate results deviate slightly, and for $n=5$ the grid spacing becomes too large and all features of the TCF are lost. In panel $\mathrm{(c)}$, we further approximate $T$ by neglecting all diagonals $\nu$, where $\nu \geq \ell$, from $DVR_{\alpha}$. We see that the approximate results lie directly on top of the exact results up to $\ell = 16$. In other words, $87.5\%$ of the diagonals in $DVR_{\alpha}$ can be neglected without impacting the results of the calculation. This is important because each additionally neglected diagonal results in a shallower circuit. Finally, we see in panel (d) that our method of computing the TEO matrix elements is robust to the cumulative error resulting from varying multiple parameters simultaneously. These results are encouraging and demonstrate that evolution under the kinetic energy operator can be well approximated by directly exponentiating $DVR_{\alpha}$, and that the Trotter error inherent in Eq.~(\ref{final-dvr}) and introduced through PITE is manageable for $N=40$. Further, only $6$ qubits and $l=4$ diagonals are needed to achieve sufficient accuracy to simulate a proton transfer model.

In this initial paper, we propose and develop the theory of hPIMC. We show that hPIMC is ideally suited for condensed phase systems and enables the study of higher-dimensional systems that are currently inaccessible through classical algorithms. We demonstrate that the PITE algorithm is ideally suited for computing short-time imaginary time evolution. We also proposed a quantum algorithm for computing evolution under the kinetic energy operator in position space using a DVR approximation and demonstrated its practicality by computing the TCF of a proton transfer reaction. How hPIMC performs in practice is a future direction that we will pursue.
Finally, the work described here represents a new direction in solving NISQ era quantum dynamics problems.

\begin{acknowledgments}
The authors acknowledge funding for this work through
NSF EAGER: QCA-QSA grant number CHE-2038005.\\
The authors thank Peter McMahon for insightful
discussion and guidance.
\end{acknowledgments}
\bibliography{apssamp}
\end{document}